# Indirect Interactions Between Proton Donors Separated by Several Hydrogen Bonds

Takaya Ogawa,[1] Hidenori Ohashi,[2] Takanori Tamaki,[3,4] Takeo Yamaguchi[3,4]

[1]*Department of Chemistry, Massachusetts Institute of Technology, 77 Massachusetts Avenue, Massachusetts 02139, USA*
[2]*Department of Chemical Engineering, Tokyo University of Agriculture and Technology, 2-24-16 Naka-cho, Koganei, Tokyo 184-8588, Japan*
[3]*Laboratory for Chemistry and Life Science, Institute of Innovative Research, Tokyo Institute of Technology, Nagatsuta 4259, Midori-ku, Yokohama 226-8503, Japan*
[4]*Kanagawa Academy of Science and Technology, 4259 Nagatsuta, Midori-ku, Yokohama 226-8503, Japan*

**Abstract**
We expand the definition of our recently proposed proton conduction mechanism, the packed-acid mechanism, which occurs under conditions of concentrated proton donors. The original definition stated that acid-acid interactions, which help overcome the barrier of the rate-determining step, occur only when a hydrogen bond is formed directly between proton donors. Here, it is shown that proton donors can interact with each other even when the donors are separated via several H-bonds. The effect of these interactions on proton diffusivity is confirmed by *ab initio* calculations.

**Main text**

Aqueous proton conduction is a versatile phenomenon in acid-base chemistry and most biological redox reactions. The generally accepted proton conduction mechanism is the Grotthuss mechanism, in which protons move along a hydrogen bond (H-bond) network through sequential hoppings and reorientations of excess protons from one solvating water molecule to the next.[1-3] Although Raman and infrared spectroscopy investigations,[4-6] X-ray spectroscopy and scattering measurements,[7-9] and computational simulations[10-13] have progressed our understanding of the behavior of excess protons in bulk water, the details of the mechanism are still elusive. The Grotthuss mechanism involves necessary co-ordination constructed by hopping, in which a proton hops from a proton donor ($P_D$; a molecule that can provide a proton to other molecules) to a proton acceptor ($P_A$; a molecule that can accept a proton from another molecule) along a H-bond, and by reorientation, in which a H-bond is cleaved and the proton reorients to another $P_A$.[14,15] The rate-determining step is reorientation because the positive charge of a proton generates a H-bond that is too strong to break.[16,17] Proton conduction generally requires many water molecules for reorientation. Water molecules with protons form Zundel ($H_5O_2^+$) and Eigen ($H_9O_4^+$) cations, and reorientation occurs at the second hydration shell, which is prompted by the thermal fluctuations of water molecules.[17-19] Protons are conducted as water-clusters with successive hopping and reorientation. This conduction mechanism has been referred to as structural diffusion and identified as the Grotthuss mechanism until recent times.[20,21] The movement of water molecules is thus indispensable for this mechanism, so that protons are not generally conducted through this mechanism when water does not move in the system.[22-24] Theoretical research has only focused on the interaction between one proton and water molecules with models that include only one excess proton and many water molecules in a unit cell, *i.e.*, diluted conditions. However, there have only been a few reports on the interactions between excess protons under concentrated conditions.[25]

Research has revealed that the Grotthuss mechanism in bulk water can be subdivided into two mechanisms.[22,25] One mechanism is structural diffusion, which occurs in the case that several $P_D$s exist at a distance and have no significant interaction with each other (diluted conditions). The second mechanism is a packed-acid mechanism, which occurs when

P$_D$S exist in close proximity and have an effect on each other (concentrated conditions). In the packed-acid mechanism, the interactions between P$_D$S, *i.e.*, the acid-acid interactions, weaken the strong H-bonds formed due to the positive charge of the proton and facilitates reorientation at H-bonds in the first solvation shell (H-bond$^{1st}$).[25] A simple way to distinguish the two mechanisms is to focus on the H-bonds in the solvation shell where reorientation occurs. Reorientation occurs at H-bond$^{1st}$s in the packed-acid mechanism and at H-bonds in the second hydration shell in structural diffusion. The packed-acid mechanism can facilitate proton conduction through acid-acid interactions without water fluctuation, and thus protons can move through frozen water.[22,25] Proton conduction in frozen water is a phenomenon that occurs in proton pumps in the human body,[26,27] which still involves unclear mechanisms such as proton selectivity.[27,28] Analysis of the packed-acid mechanism is expected to provide insights into proton diffusion in biological systems. Proton conducting materials for applications such as fuel cells typically involve proton movement via structural diffusion.[15,22] Structural diffusion requires high humidity to retain the high proton diffusivity, which requires water management systems that negatively affect the energy efficiency and increase costs.[29,30] In contrast, the packed-acid mechanism does not require water movement, which results in high proton diffusivity without the need for water management.[22] Despite the importance for science and industry, investigation of the packed-acid mechanism is not yet sufficient.

The original packed-acid mechanism was proposed as the mechanism for proton conduction via direct H-bonds between P$_D$S. In this study, we theoretically show that P$_D$S can indirectly affect each other despite being separated by several H-bonds (indirect acid-acid interactions), which facilitates reorientation of the H-bond$^{1st}$. Thus, the definition of the packed-acid mechanism is extended to include indirect acid−acid interactions. First, the source of the interactions is explained qualitatively. The quantitative influence of such interactions on the H-bond network is then confirmed by density functional theory (DFT) calculations. Consequently, such indirect interactions are confirmed as a certain contribution to proton conduction.

To explain this concept, the following four types of H-bond networks are presented: P$_D$S in H-bond networks separated distantly (Fig. 1a, only one P$_D$ is described), moderately (Fig. 1b), closely (Fig. 1c), and not indirectly (Fig. 1d). The four types of H-bond network differently influences on reorientation. The quantitative distance between P$_D$S was evaluated with respect to the way point number (*WPN*; see Fig. S1 for detailed explanation), which is defined as the minimum number of way points (P$_A$S) between two P$_D$S through H-bonds because the H-bond is the important factor for indirect acid-acid interactions (H-bond network in Figs. 1a, b, c, and d are described as networks with *WPN* =∞, 3, 1, and 0, respectively).

The general effect of P$_D$ is illustrated in Fig. 1a. A mobile proton exists in P$_D$ molecules, and the proton tends to intramolecularly polarize to positive. The positive charge of the proton in P$_D^{1+}$ extracts electron density from P$_A^{0\pm}$ and forms a too rigid H-bond (blue double circled H-bond in Fig. 1a), which is unsuitable for reorientation (the rate-determining step). This phenomenon accompanies intermolecular polarization where the electron density of P$_A$ shifts to P$_D$, which results in P$_D^{(1-\delta)+}$ and P$_A^{\delta+}$ (see Fig. S2). The positively polarized P$_A^{\delta+}$ behaves similar to a P$_D$ that can extract electron density from another P$_A$ to form another rigid H-bond (green double circled H-bond in Fig. 1a). This phenomenon can happen successively through H-bonds. Therefore, polarization occurs through some H-bonds and forms a rigid H-bond network.[31,32]

In contrast, the packed-acid mechanism is depicted in Fig. 1d as a case where the packed-acid mechanism was derived from direct acid-acid interaction. P$_D$ does not receive an extra proton and rarely forms a H-bond with another P$_D$ because of its low proton acceptability.[25] The H-bond between P$_D$S then becomes easy to break, which results in reorientation of the H-bond$^{1st}$.

In the H-bond network, in which P$_D$S are separated moderately and closely, an electron split occurs in the center P$_A$ (the green wavy lined water molecule in Figs. 1b and c). In Fig. 1c, the electron split is where the electron density on P$_A$ is extracted by two P$_D$S, and the electron density per H-bond is divided. Therefore, the electron density of P$_A$ to form a H-bond decreases, and both H-bonds become weak (red double circled H-bonds in Fig. 1c, see Fig. S3). Thus, two $P_{D}s$ can indirectly weaken the H-bond$^{1st}$ shell through one water molecule, *i.e.*, indirect acid-acid interactions.

Furthermore, an electron split can influence H-bond$^{1st}$ through more than one $P_A$. As shown in Fig. 1a, a $P_A$ that is affected by a $P_D$ though some H-bonds can behave similarly to a $P_D$. An electron split must thus occur indirectly by two H-bonds from the $P_D$-like $P_{A}s$ (the green wavy lined water molecule in Fig. 1b). Therefore, these H-bonds are weakened (grey double circled H-bonds in Fig. 1b) and the weakened H-bonds induce low polarization of the $P_{A}s$ (see Figs. S4 and S5). Therefore, the $P_D$-like $P_A$ does not polarize sufficiently to accept a H-bond from another molecule. The successive H-bonds are thus weakened, which leads to successive low polarizations. The H-bond$^{1st}$ is eventually weakened (red double circled H-bonds in Fig. 1b), *i.e.*, indirect acid-acid interactions. Therefore, reorientation can occur in H-bond$^{1st}$ by indirect acid-acid interactions as well as by direct acid-acid interactions, as shown in Fig. 1d.

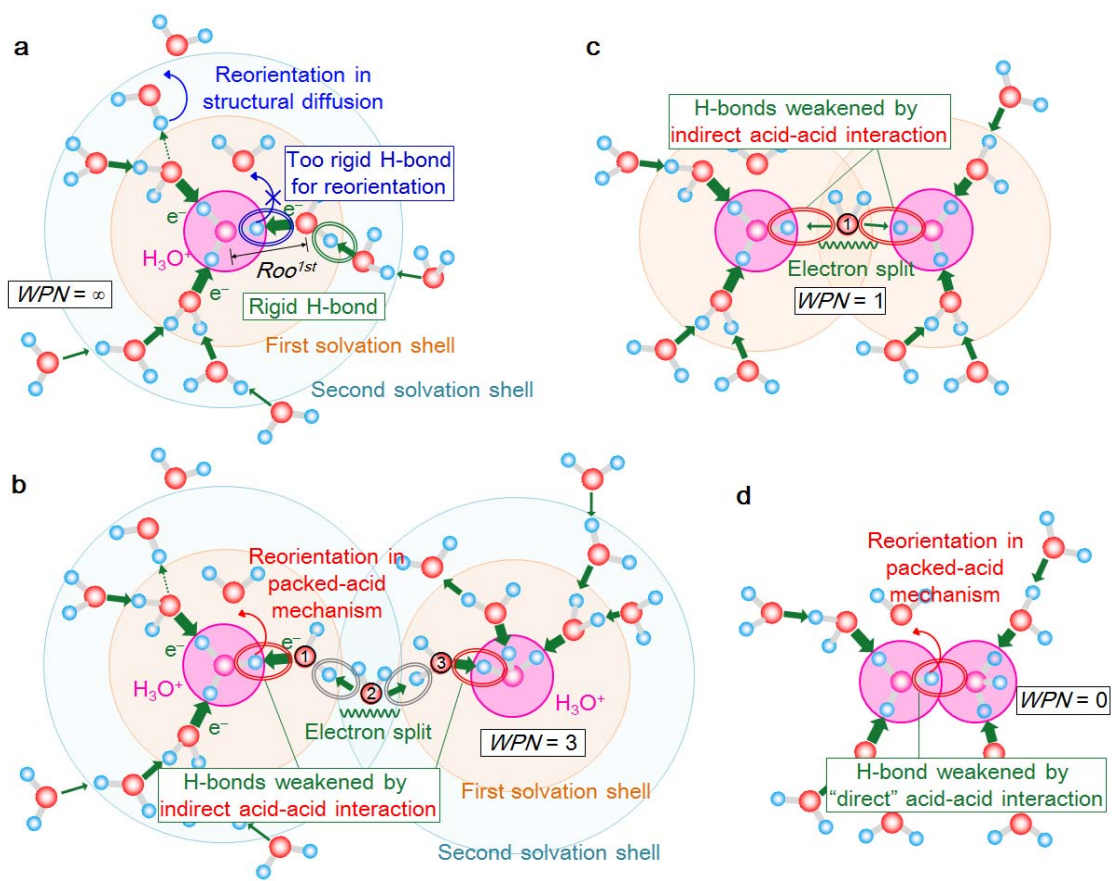

Figure 1. **a** Normal hydrogen bond (H-bond) network comprised of water molecules around a proton donor ($P_D$). This illustration also depicts the oxygen-oxygen distance of H-bonds in the first solvation shell ($R_{OO}^{1st}$) and images of reorientations during structural diffusion in the first and second hydration shells. $P_D$ does not interact with other $P_{D}s$ ($WPN = \infty$) **b** H-bond network generated by two $P_{D}s$ that are moderately close to one another and some water molecules with $WPN = 3$. This illustration also shows reorientation during the packed-acid mechanism. **c** H-bond network generated by two $P_{D}s$ in much closer proximity with $WPN = 1$. **d** H-bond network including direct H-bond between two $P_{D}s$ ($WPN = 0$).

Based on this concept, we propose that proton conduction can occur successively via indirect acid-acid interactions. Figure 2 shows schematic illustrations of the mechanism, where it is assumed that a proton conducts from left to right through the ordered water molecules in the center. In Fig. 2a, the pink-circled $H_3O^+$ interacts with the blue-circled $H_3O^+$ through $s$ ($WPN = s$) water molecules and generates an indirect acid-acid interaction. The interaction weakens the red-circled hydrogen bond and reorientation occurs. In Fig. 2b, the reoriented bond can emit a proton to the next water molecule in the first solvation shell, although the elongated bond by the indirect acid-acid interaction ($WPN = t$) slightly disturbs the hopping, which is not the rate-determining step, as discussed in following section. In Fig. 2c, an indirect acid-acid interaction occurs between blue- and pink-circled $H_3O^+$ ions through $u$ water molecules ($WPN = u$), and reorientation is facilitated in the same way as Fig. 2a. The subsequent hopping can also occur in the same way as shown in Fig. 2b. Therefore, indirect acid-acid interactions can facilitate proton conduction successively. It is noted that hopping and reorientation need not occur in the same *WPN*.

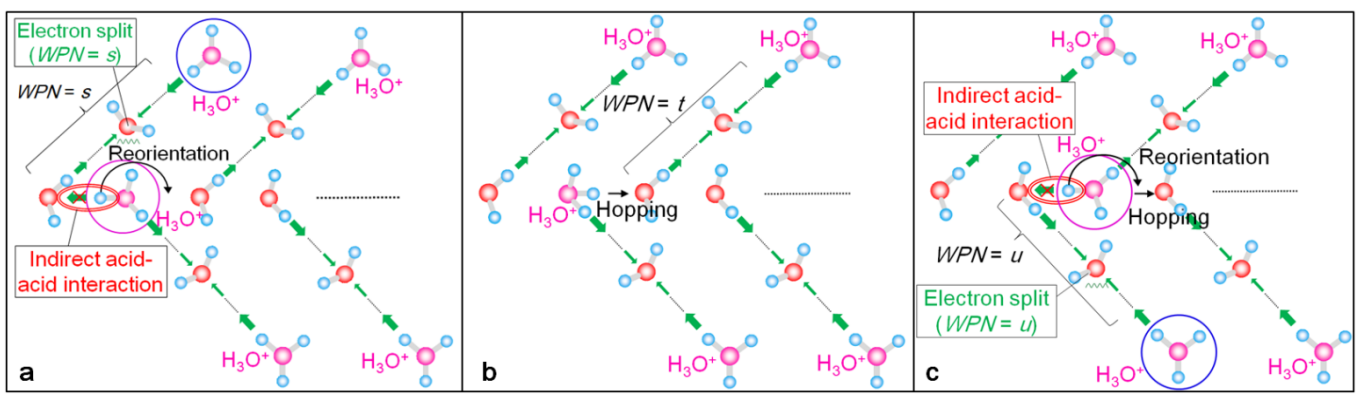

Figure 2. Concept of proton conduction facilitated by indirect acid-acid interactions (see text).

The next question is to what extent do indirect acid-acid interactions influence proton diffusivity because the intermolecular polarization due to H-bond to generate $P_D$-like $P_{AS}$ weakens as the distance from $P_D$ increases.[16] In order to discuss the influence, DFT calculations were performed using optimized models composed of two extra protons and water molecules with a periodic boundary condition. The number of water molecules per proton was varied from two to fifteen. The sizes of the unit cells were determined to describe the density as 1 g/cm$^3$ (details of calculation methods and models are given in the Supplemental Material). Proton diffusivity is described as $D = D_0/T \exp(-E_a/RT)$, where $E_a$ is the activation energy, R is the gas constant, $T$ is the temperature, and $D_0$ is a prefactor. $E_{a:x}$ ($x$ = hopping, reorientation) can be approximated as a function of the oxygen-oxygen distance for a H-bond$^{1st}$ ($R_{OO}^{1st}$) (see Figs. S6, S7, and S8 for detailed derivations):

$E_{a:hopping} = 381.21r^2 - 1783r + 2083.4$   $E_{a:reorientation} = 1/(-0.0588r^2 + 0.4303r - 0.6857)$, where $r = Roo^{1st}$ (Å)   (1)

Therefore, the contribution of indirect acid-acid interactions to proton diffusion is evaluated with respect to how $E_{a:reorientation}$ of H-bond$^{1st}$ ($WPN \geq 1$) decreases due to elongated (weakened) $R_{OO}^{1st}$. $E_{a:hopping}$ was also investigated because there is a possibility that a too long H-bond$^{1st}$ would make hopping the rate-determining step. The step where H-bond$^{1st}$ expands and contracts exists between hopping and reorientation. However, the step is so fast at experimentally reasonable temperatures for a liquid that it does not become the rate-determining step (see section V in Supplemental Material).

Here, the contribution of indirect acid-acid interactions ($WPN \geq 1$) to proton diffusivity is estimated with respect to direct acid-acid interactions ($WPN = 0$). To discuss the influence of indirect acid-acid interactions based on WPN, four parameters were defined as follows: $M(WPN)$, $\rho(WPN, Roo^{1st})$, $Q_x(WPN)$, and $MQ_x(WPN)$. $M(WPN)$ is the proportion of H-bond$^{1st}$ that belongs to each WPN. In the model investigated, $M(0)$, $M(1)$, $M(2)$, $M(3)$, $M(4)$, and $M(5)$ were 2%, 30%, 41%, 23%, 5%, and 1%, respectively (Fig. 3a). The H-bond$^{1st}$ group associated with each $WPN$ has typical distributions of $Roo_{1st}$, which can be described as the radial distribution function of $Roo_{1st}$, defined as $\rho(WPN, Roo_{1st})$ (Fig. 3b). As $WPN$ decreases (a short distance between two $P_D$s,), $\rho(WPN, Roo_{1st})$ gradually shifts to the region of long $Roo_{1st}$ (strong indirect acid-acid interactions). The shift of $\rho(WPN, Roo^{1st})$ was quantitatively evaluated using $Q_x(WPN)$:

$$Q_x(WPN) = \int \rho(WPN, Roo^{1st}) \exp\left(-\frac{E_{a:x}(Roo^{1st})}{RT}\right) dRoo^{1st} \quad (x = hopping, reorientation) \quad (2)$$

Here, $Q_x(WPN)$ represents the value where $\rho(WPN, Roo^{1st})$ is weighted by $E_a$. The sum of $MQ_x(WPN)$ (= $M(WPN) \times Q_x(WPN)$ ($x$ = hopping or reorientation, where $x$ is the rate-determining step)) is proportional to the actual proton diffusivity and the index to evaluate the contribution to proton diffusion. $Q_x(WPN)$ and $MQ_x(WPN)$ at room temperature (298.15 K) are shown in Figs. 3c and 3d, respectively. Hopping in the $WPN = 0$ model means hopping from $H_3O^+$ to $H_3O^+$, which results in $H_2O$ and $H_4O_2^{2+}$. Because this phenomenon should rarely occur, the hopping in WPN = 0 was excluded. In Fig. 3c, all $Q_{hopping}(WPN)$ are higher than all $Q_{reorientation}(WPN)$, which is the same tendency as with $MQ_x(WPN)$ (Fig. 3d). These results indicate that reorientation is still the rate-determining step, even though $R_{OO}^{1st}$ is elongated by indirect acid-acid interactions. Thus, only reorientation was investigated in the following discussion.

The shift of $\rho(WPN, Roo_{1st})$ in Fig. 3b induces a result where $Q_{reorientation}(0)$ is larger than $Q_{reorientation}(WPN \geq 1)$. However, as shown in Fig. 3a, the proportion of H-bonds$^{1st}$ group of $WPN = 0$, $M(0)$ is very small in our models. Figure 3e shows that the sum of $MQ_{reorientation}(WPN \geq 1)$ is larger than $MQ_{reorientation}(0)$. Therefore, indirect acid-acid interactions ($WPN \geq 1$) further contribute to proton diffusivity more than direct acid-acid interactions ($WPN = 0$). In addition, the proportional values of $MQ_{reorientation}(0)$, $MQ_{reorientation}(1)$, $MQ_{reorientation}(2)$, $MQ_{reorientation}(3)$, and $MQ_{reorientation}(4)$ against the sum of $MQ_{reorientation}(WPN \geq 0)$ were 37.1%, 42.3%, 15.6%, 4.8%, and 0.2%, respectively. Thus, $MQ_{reorientation}(4)$ is very small, and should be regarded as negligible in all proton diffusion in the model used here. Hence, indirect acid-acid interactions contribute to proton diffusion in the H-bond networks where $P_D$s exists within close proximity, *i.e.*, a situation with packed-acids.

Acid-acid interactions (packed-acid mechanism) occur through several H-bonds. The definition of $P_D$ includes not only $H_3O^+$ but $SO_3H$ and $PO_3H_2$, and moreover, the definition of $P_A$ includes not only $H_2O$ but also $SO_3^-$ and $PO_3H^-$. Therefore, acid-acid interaction can occur even in the solid state with functionalized acid-base groups. $\rho(WPN, Roo_{1st})$, $M(WPN)$, and the optimal $WPN$ to cause indirect acid-acid interactions must change depending on the $P_D$s and $P_A$s. In addition, $M(WPN)$ should be controllable in the solid state because groups can be functionalized to achieve regulated distances. Therefore, electrolytes in which protons are conducted via the packed-acid mechanism can be designed so that proton conduction that is less dependent on relative humidity can be derived, which is very important for numerous applications. Moreover, indirect acid-acid interactions can occur in biological systems as proton conduction without water movement. The meaning of the acid-base allocations in proton paths within proteins may be explained by acid-acid interactions. The findings presented here will contribute to several research areas ranging from engineering to science.

In this communication, although the focus was on H-bond$^{1st}$s weakened by acid-acid interactions, H-bonds in other solvation shells are also elongated through acid-acid interactions. Future investigations that address the reorientations at H-bonds in other solvation shells are currently underway.

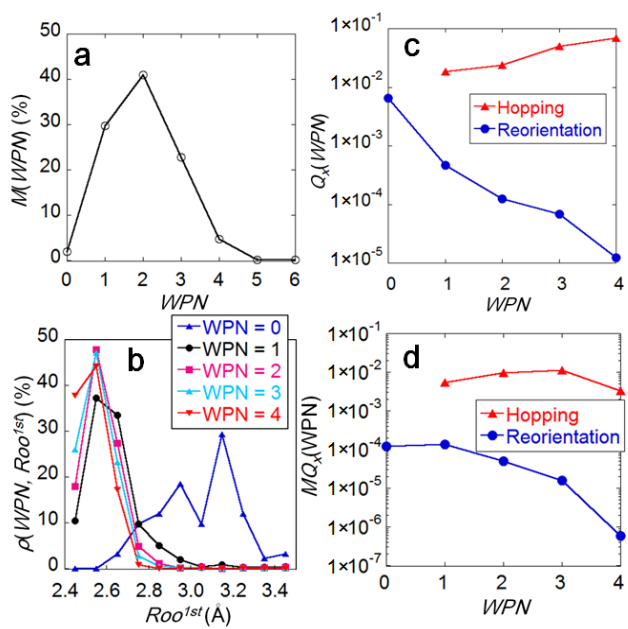

Figure 3. **a** $M(WPN)$ as a function of $WPN$. **b** $\rho(WPN, Roo^{1st})$ in bulk water with protons as a function of $Roo^{1st}$ for various $WPN$. The number of H-bond$^{1st}$ was counted by 0.1 Å from 2.4 Å to 3.5 Å, and divided by the total number of H-bond$^{1st}$. The middle value of $Roo^{1st}$ in each interval represents the value in the x-axis (e.g. 2.45 Å represents the ratio of H-bond$^{1st}$ in which $Roo^{1st}$ is between 2.40 and 2.50 Å). **c** $Q_x(WPN)$ as a function of $WPN$. **d** $MQ_x(WPN)$ as a function of $WPN$.

**Acknowledgments**

This work was supported by the Kanagawa Academy of Science and Technology and Grant-in-Aid for JSPS Fellows.

# Supplemental Material

# Indirect Interaction Between Proton Donors Separated by Several Hydrogen Bonds

Takaya Ogawa,[1] Hidenori Ohashi,[2] Takanori Tamaki,[3,4] Takeo Yamaguchi[3,4]

[1]Department of Chemistry, Massachusetts Institute of Technology, 77 Massachusetts Avenue, Massachusetts 02139, USA

[2]Department of Chemical Engineering, Tokyo University of Agriculture and Technology, 2-24-16 Naka-cho, Koganei, Tokyo 184-8588, Japan

[3]Laboratory for Chemistry and Life Science, Institute of Innovative Research, Tokyo Institute of Technology, Nagatsuta 4259, Midori-ku, Yokohama 226-8503, Japan

[4]Kanagawa Academy of Science and Technology, 4259 Nagatsuta, Midori-ku, Yokohama 226-8503, Japan

**Abbreviations**

$P_A$: proton donor, a molecule that can provide a proton to other molecules

$P_D$: proton acceptor, a molecule that can accept a proton from another molecule

H-bond: hydrogen bond

H-bond$^{1st}$: H-bond in the first solvation shell

WPN: way point number, the minimum number of way points ($P_A$s) between two $P_D$s through H-bonds

DFT: density functional theory

*Roo*: the length of the oxygen-oxygen distance along H-bond

*Roo*$^{1st}$: the length of the oxygen-oxygen distance along H-bond$^{1st}$

$E_a$: activation energy

## I. Detailed definition of *WPN*

Figure S1 shows one example of a bulk water system with two excess protons, which is investigated in the main text. The structures were visualized using VESTA (Visualization for Electronic and STructural Analysis), where blue and red atoms are respectively hydrogen and oxygen. In the main text, the definition of *WPN* is the minimum number of way points ($P_{AS}$) between two $P_{DS}$ through H-bonds. The reason that minimum was added in the definition is because $P_{DS}$ usually interact via several paths of H-bonds, as shown in Fig S1. The example in Fig. S1 shows two H-bonds paths along arrows between two pink-circled $H_3O^+$ ions that interact with each other, and electron split occurs at the blue-circled $H_2O$. The H-bond path along red arrows is composed of the minimum number of $P_{AS}$ between the $P_{DS}$ (*WPN* = 2). If the H-bond path bypassed along black arrows, then the number of $P_{AS}$ is four. However, the H-bond path with the minimum number of $P_{AS}$ should be dominant because the force of interaction decreases by passing through $P_{AS}$, according to Ref. 31 (Table 2 in Ref. 31 suggests that $P_D$-like $P_A$ does not have the same ability as $P_D$ to make a rigid H-bond). Therefore, it is assumed that *WPN* can simply describe the interaction between $P_{AS}$. The validity of this assumption was confirmed by $\rho$(*WPN*, $Roo_{1st}$) in Fig. 3b, which shows that the distribution gradually shifts from the short to the long $Roo^{1st}$ region as *WPN* decreases one by one.

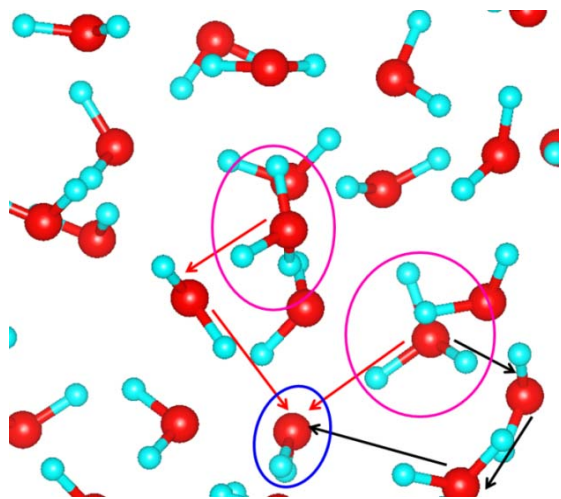

Figure S1. One example of bulk water with two excess protons.

## II. Detailed information on DFT calculations

For the theoretical analysis, we performed density functional theory (DFT) calculations using the revised Perdew–Burke–Ernzerhof functional. Single Kleinman–Bylander projectors were used to represent each angular momentum channel and a relativistic polarized calculation was used to estimate the effect of spin. A double-zeta split-valence basis set with polarization orbitals (DZP) was used. Norm-conserving pseudopotentials were applied using the improved Troullier–Martins method with nonlinear core corrections. All calculations were performed using the SIESTA (Spanish Initiative for Electronic Simulations with Thousands of Atoms) 2.0.1 software package.

The models consist of several water molecules and two excess protons with a periodic boundary condition and $2 \times 2 \times 2$ k-grid sampling. The net charge of models is +2.0. The form of the unit cell is cubic and the lattice constant for models are determined to maintain the density of water molecules at 1 g/cm$^3$. The numbers of water molecules in the unit cells were 4, 6, 8, 10, 20, and 30, *i.e.*, the concentrations of protons were 27.8, 18.5, 13.9, 11.1, 5.6, and 3.7 mol/L, respectively. Each model was constructed with water molecules allocated randomly in the unit cell. The patterns of random allocation were over 60 per model with different numbers of water molecules in the unit cell, *i.e.*, over 360 patterns of models in total. *Ab initio* molecular dynamics simulations were performed for the models at 363 K for 0.5 ps using a Nosé thermostat to control the system temperature. After these allocations, the models were optimized with the termination condition that the maximum tolerance force was 0.01 eV/Å.

An H-bond was defined as follows: bond angle > 110° according to IUPAC and $R_{oo}$ > 2.4 Å but < 3.5 Å. The minimum $R_{oo}$ value found in these simulations was 2.4 Å. The experimental distribution function indicates that the first shell of water exists with an upper limit of 3.5 Å. Therefore, the maximum $R_{oo}$ value was set to 3.5 Å. It was assumed that one proton can belong to only one H-bond. If one proton had two possible candidates, then the proton was assigned to the H-bond with the shorter $R_{oo}$ value. The definition was employed to estimate *WPN*.

In following sections, partial atomic electrostatic charges were computed with the Mulliken scheme. To calculate the Mulliken charge of each atom, the basis set of hydrogen, DZP, was changed to the double-ζ split-valence basis set (DZ) while fixing the structure calculated with the DZP basis set. In the case of a single water molecule, the DZP basis set for the hydrogen atom including p-type orbitals gave the electron population number of the p-type orbital of the hydrogen atom as 0.178 and the Mulliken atomic charge of protons becomes a strange value, –0.052. On the other hand, the Mulliken charge of protons with DZ for hydrogen and DZP for oxygen atoms while the optimized geometry was fixed with the DZP basis set was +0.230 in the case of a single water molecule, which is a reasonable value. Therefore, in this Supplementary Material, the optimized structures were determined using the DZP basis set for all atoms and the Mulliken charge was then calculated using the DZ basis set for hydrogen and the DZP basis set for all other atoms. It should be noted that this observation of the Mulliken atomic charge of protons calculated with too many polarization orbitals is also a feature of the quantum chemistry calculation with the linear combination of atomic orbitals (LCAO) approximation.

### III. Changes in electron density and hydrogen bond length by indirect acid-acid interactions

DFT calculations were performed to qualitatively clarify indirect acid-acid interaction using the optimized model consisting of 2−3 water molecules. The calculation method is same as that shown in section I. The unit cell is a cube with a lattice constant of 30 Å to avoid artificial interactions between unit cells.

First, a dimer of water molecules is shown in Fig. S2, which describes $R_{oo}$ between two water molecules and the partial charge of the atoms. The total net charges of the blue- and red-circled water molecules are –0.070 and +0.070, respectively. Therefore, for a H-bond between two molecules, the electron density is extracted by the molecule that has a proton to form a H-bond from the molecule that receives the H-bond. This occurs in the H-bond between a $P_D$ and a $P_A$, which results in $P_D^{(1-\delta)+}$ and $P_A^{\delta+}$.

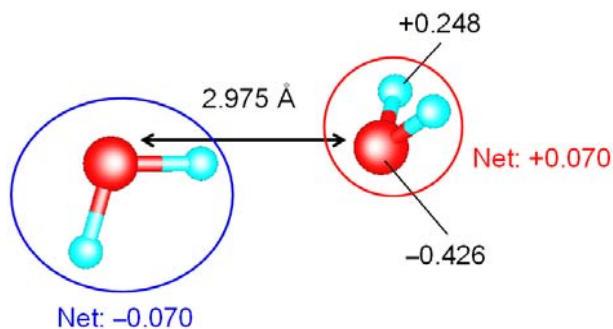

Figure S2. Dimer of water molecules.

Electron split occurs when one water molecule receives two H-bonds, as shown in Fig. S3. One water molecule was added to the dimer model to form a H-bond. The charge of oxygen in the red-circled water molecule changes from –0.426 (see Fig. S2) to –0.404, due to the electron density extraction from the new H-bond. Therefore, the decreased electron density form H-bonds less strongly and $R_{oo}$ of the original H-bonds increased from 2.975 Å to 3.038 Å. This phenomenon occurs in the $P_A$ that receives H-bonds from two $P_D$s, *i.e.*, electron split, as discussed in the main text.

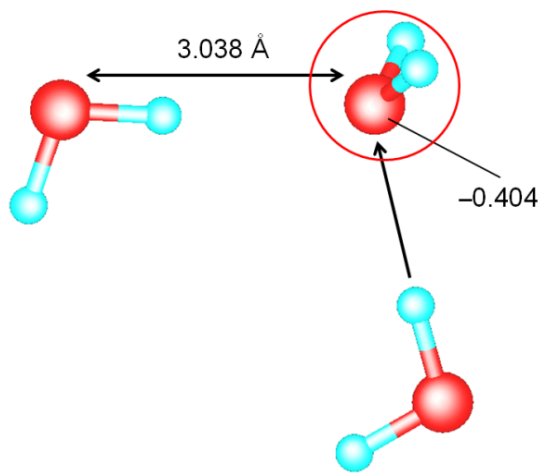

Figure S3. Example of three water molecules, where one water molecule receives two H-bonds.

A hydrogen bond becomes strong when the molecule that receives the H-bond forms another H-bond from itself, as shown in S4. One water molecule is added to the dimer model to allow the red-circled water molecule to form a H-bond. The electron density of the added water molecule is extracted to the red-circled water molecule. The net charge of the red-circled water molecule then shifts from +0.070 to +0.002, and the charge of the oxygen in the red-circled water molecule changes from –0.426 (see Fig. S2) to –0.430. Therefore, more electron density is engaged to form the original H-bond, and the $Roo$ of the original H-bond is shortened from 2.975 Å to 2.920 Å. Therefore, when $P_D$-like $P_A$ (see main text) forms a H-bond with another $P_A$ from $P_D$-like $P_A$, the H-bond between $P_D$ and $P_A$ becomes strong. Conversely, if $P_D$-like $P_A$ cannot form a H-bond from $P_A$ itself, the H-bond between $P_D$ and $P_A$ becomes weak. Therefore, the electron split weakens the H-bond from $P_D$-like $P_A$, and the weakened H-bond induces a weak H-bond between $P_A$ and $P_D$. This phenomenon can occur in the case of a H-bond that is separated from $P_D$ by some H-bonds, *i.e.*, between $P_D$-like $P_A$S. Consequently, the electron split can have an influence on H-bond$^{1st}$.

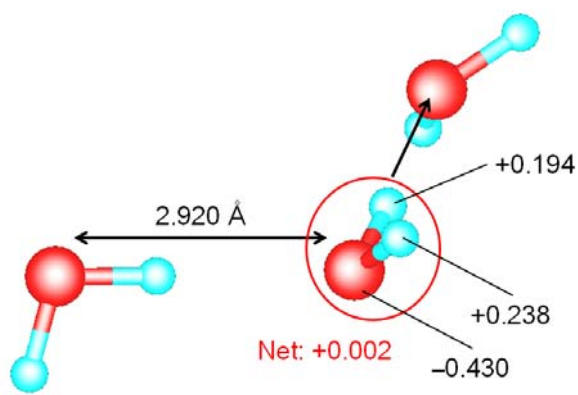

Figure S4. Example of three water molecules, where the water molecules successively form H-bonds in the same direction.

We now focus on the case where an electron split occurs, as shown in Fig. S5a, and the H-bond along the red arrow becomes weak. However, the weakened H-bond$^{1st}$ appears even when an electron split does not occur. Figure S5b shows the case where $P_D$-like $P_A$ forms a H-bond from $P_A$ with another $P_D$. The proton affinity of $P_D$ is very low according to Ref. 25, so that $P_D$-like $P_A$ cannot form a rigid H-bond. For the same reason associated with the example in Fig. S4, the H-bond along the red arrow in Fig. 5b is also weak. Eventually, indirect acid-acid interaction also occurs in the case shown in Fig. S5b.

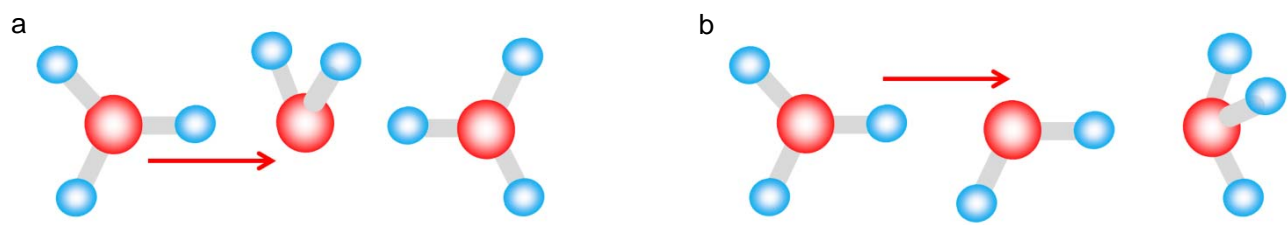

Figure S5. Concept of indirect acid-acid interaction **a** with and **b** without an electron split.

## IV. Equation of $E_a$ for hopping and reorientation as a function of $Roo^{1st}$

Equations of $E_a$ for hopping and reorientation as a function of $Roo^{1st}$ in the main text were derived from the data in Ref. 25. Figure S6 shows all plots of $E_a$ for hopping and reorientation as a function of $Roo$. Although the plots contain large scatter, valid points for proton conduction can be observed in the bottom of the distribution. Both plots were converted into $I_x$ distributions, which is defined as:

$$I_x = \exp\left(-\frac{E_{a:x}}{RT}\right) \quad (x = hopping, reorientation) \quad (3)$$

where R is the gas constant and $T$ is the temperature. Figure S7 shows $I_x$ as a function of $Roo$ at 298.15 K. Almost all of the points that are dispersed in the upper regions of Figs. S6(a) and (b) are suppressed by the exponentiation, as shown in Fig. S7. Therefore, we consider that the points at the bottom of the distribution are dominant for proton conduction.

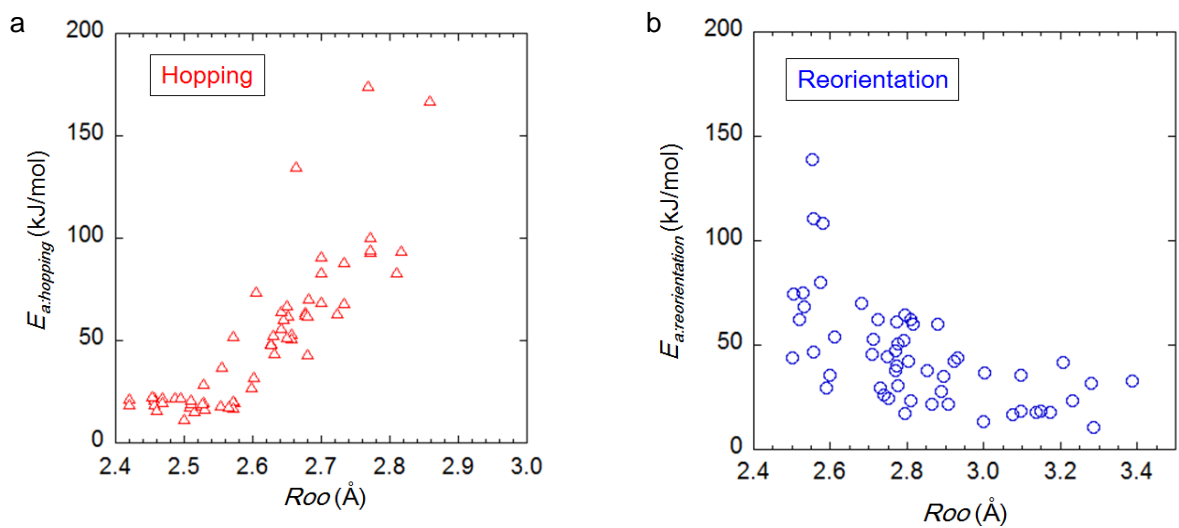

Figure S6. Relation between $Roo$ and $E_a$ for **a** hopping and **b** reorientation. Panels were adapted from Ref. 25 with permission from the Royal Society of Chemistry (Copyright 2014) and are repeated here for clarity.

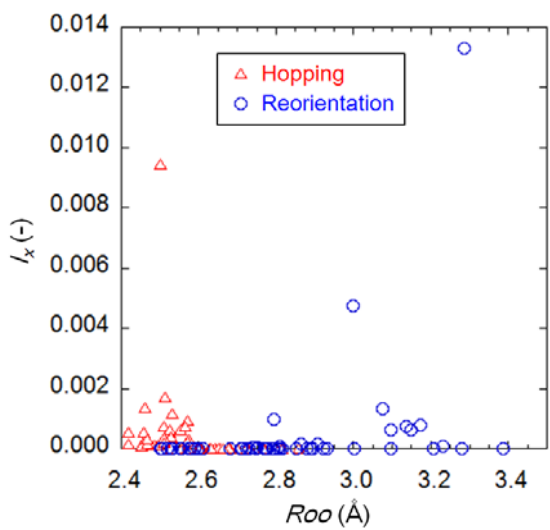

Figure S7. Plots of $I_x$ as a function of $Roo$.

We selected some points along the bottom of the distribution to derive the fitting equations. The energy diagram for hopping is described as a double potential, which suggests that the motion of protons is approximated as a spring vibration. Therefore, the equation for $E_{a:hopping}$ should be a quadratic function. A quadratic function was fitted to the selected plot using a least mean squares method (Fig. S8a with the fitting equation). In these calculations, for a H-bond from $H_3O^+$ to $H_2O$ with $Roo$ = 2.4 Å, $H_5O_2^+$ is formed and the proton is present almost at the center between the two oxygens. Thus, the double potential becomes a single potential and $E_{a:hopping}$ is almost zero. A point, ($Roo$, $E_a$) = (2.4, 0) was then included for fitting; however, the equation did not change significantly. The equation fits the points well and the coefficient of determination ($R^2$) is almost 1.0000.

$E_{a:reorientation}$ is based on the binding energy of the H-bond. Using the same idea as with the case of hopping, the energy should be described as a spring formula:

$$E_{a:reorientation} = A - B(r + C)^2 \quad (4)$$

where $r$ is $Roo$, and A, B and C are constants. However, the equation does not fit with the selected plots and $R^2$ = 0.9652. This may suggest that the approximation as a spring is too rough because $Roo$ of the plots for reorientation was too long at 2.5–3.3 Å, whereas $Roo$ of the plots for hopping were within *ca.* 2.8 Å. Therefore, instead of equation (4), an inverse quadratic function was employed, which fitted well with the points (Fig. S8b with the fitting equation). These equations can be applied to $Roo^{1st}$. $Roo$ at the intersection point of two equations is 2.61 Å.

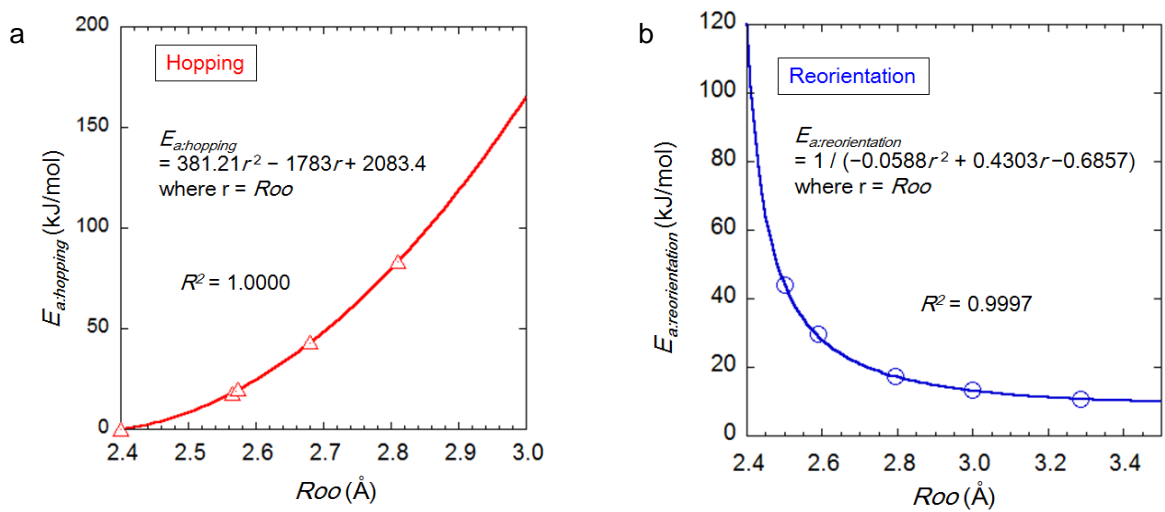

Figure S8. Selected points for fitting to derive the equations of **a** $E_{a:hopping}$ and **b** $E_{a:reorientation}$. The equations for each procedure are also shown.

## V. Fast process in which H-bond[1st] expands and contracts

Between the hopping and reorientation events, there is a process in which H-bond[1st] expands and contracts. The average change of $Roo^{1st}$ was estimated to clarify that this process is not the rate-determining step. The diffusivity of water molecules in bulk water is $2.3\times10^{-9}$ m$^2$/s at 25 °C, which can be considered as the movement of oxygen. The time scale for reorientation is approximately 1.5 ps according to experimental measurements.[1] $Roo^{1st}$ is considered as a one-dimensional length between two oxygens of water molecules. Therefore, the average displacement can be calculated using the square root of the mean-square displacement of one dimension, $\Delta \overline{Roo^{1st}} = \sqrt{2 \times 2Dt} = 1.17$ Å. $2Dt$ is multiplied by two in the square root term because the displacement occurs as the movement of two oxygens. The definition of H-bond is 2.4 Å < $Roo$ < 3.5 Å as stated in section II of this Supplemental Material. The length range is 1.1 Å, which is covered by the average displacement of *ca*. 1.2 Å. Therefore, the process in which H-bond[1st] expands and contracts is not the rate-determining step for proton conduction.